# The *infinity-loop microresonator*: A new integrated photonic structure working on an exceptional surface

Riccardo Franchi ; Stefano Biasi; Diego Piciocchi; ... et. al

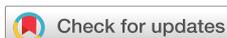 Check for updates

View Online 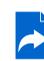 Export Citation  CrossMark

**Articles You May Be Interested In**

Angle dependent interlayer magnetoresistance in multilayer graphene stacks

*Journal of Applied Physics* (October 2015)

Determining whether diabolical singularities limit the accuracy of molecular property based diabatic representations: The $1,2\,^1A$ states of methylamine

*J. Chem. Phys.* (October 2018)

Manifestation of bifurcations and diabolic points in molecular energy spectra

*J. Chem. Phys.* (February 1990)

# The *infinity-loop microresonator*: A new integrated photonic structure working on an exceptional surface

Riccardo Franchi,[a)] Stefano Biasi, Diego Piciocchi, and Lorenzo Pavesi

AFFILIATIONS
Nanoscience Laboratory, Department of Physics, University of Trento, 38123 Trento, Italy

[a)]Author to whom correspondence should be addressed: riccardo.franchi@unitn.it

## ABSTRACT

Exceptional points, where eigenvalues and eigenvectors coalesce, impact the behavior of different photonics components that show, e.g., enhanced sensing, coherent perfect absorption, unidirectional lasing, and chirality. However, only a few passive geometries have been developed that work on these points. Here, we introduce a novel non-Hermitian structure based on a microresonator shaped as the infinity symbol twice coupled to a bus waveguide: the infinity-loop microresonator. Unlike other structures working on an exceptional surface, the infinity-loop microresonator can achieve either high- or low-contrast unidirectional reflection with a negligible or identical reflection for counterpropagating light. It allows an easy walking through the Riemann sheet by simply controlling the phase of the light propagating in the bus waveguide, which makes it a tunable component to build more complex topological structures. Furthermore, the infinity-loop microresonator allows sensors that show the features of both an exceptional point device and a diabolic point device simultaneously.

## I. INTRODUCTION

Non-Hermitian systems have found increasing interest in recent years since they describe open systems such as the one of a propagating optical field in a waveguide.[1–4] Non-Hermitian degenerations in which the eigenvalues and eigenvectors of the system coalesce characterize the physics of exceptional points (EPs).[5–20] This differs from the physics of Diabolic Points (DPs) that characterize the degeneracy of Hermitian systems where the eigenvalues coalesce while the eigenvectors remain orthogonal. In addition, EPs are of practical interest for different applications, such as enhanced sensors,[2,21–25] unidirectional lasing,[26–29] laser line-width broadening,[30] chiral transmission,[26,31–35] unidirectional reflection,[36] loss-induced transmission,[37] unidirectional invisibility,[38,39] topological energy transfer,[40] and the breaking of the Lorentz reciprocity theorem.[41–43]

A suitable platform to study EPs is integrated optics because of the easy tunability of optical structures as well as the easy realization of non-Hermitian systems that are stable over time. This has led to the realization of various miniaturized optical structures working at an EP, for example, a silicon dioxide (silica) micro-toroid cavity with two silica nano-tips,[21] the Taiji microresonator,[36,44] two microresonators coupled together and with two bus waveguides,[45] a microresonator coupled to a waveguide with at one of the ends a tunable symmetric reflector.[34] Integrated optical microresonators are particularly appealing since they can be described by a two-level Hamiltonian with propagating and counterpropagating (or clockwise and counterclockwise) optical modes as eigenstates.[46] Indeed, non-symmetrical microresonators have been used to investigate characteristic phenomena arising from the degeneration of their eigenstates. The EP degeneration causes these systems to have additional properties and advantages over those of a simple symmetrical microresonator, for example, the enhanced sensing or the unidirectional reflectivity. The geometrical asymmetry allows them to work at an EP, resulting in a spectral transmission or reflection response characterized by a single peak or dip. This spectral response is a consequence of the coalescence and, thus, the degeneracy of the eigenvalues.

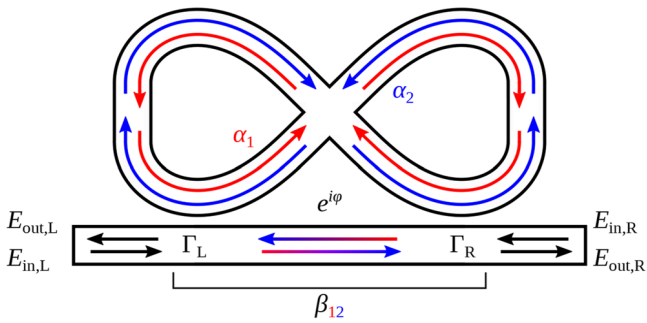

**FIG. 1.** Sketch of the Infinity-Loop Microresonator (ILMR). All the symbols are described in the main text.

In this work, we propose a new structure, called the Infinity-Loop Microresonator (ILMR), which although it works at an EP can be geometrically symmetrical and shows resonance splitting in the spectral response. The ILMR is made by a microresonator shaped like the infinity symbol and a bus waveguide coupled to both of its lobes (Fig. 1). This peculiar geometrical shape allows for preserving the spectral splitting at degeneracy and introduces other characteristic features that make this microresonator suitable for studying the physics of EP and exploiting its properties, e.g., in sensing local perturbations. In Sec. II, we model the ILMR, we demonstrate that it works at an EP, and we show that it can be easily tuned to achieve a desired spectral response. Moreover, we analyze how the ILMR behaves under perturbations of its fundamental state. We do not only compute the Riemann sheets related directly to the eigenvalues of the system,[1,3,8,10,47,48] but we also compute the Riemann sheets from the spectral responses (i.e., the actual observable of the system). In Sec. III, we compare experimental measurements of a few structures realized on the silicon photonics platform with the modeling. Here, it is shown that the introduced Temporal Coupled Mode Theory (TCMT) equations faithfully describe the experimental spectral responses and allow quantitatively extracting the relevant parameters of the ILMR. In Sec. IV, we compare the ILMR with different types of integrated microresonators, emphasizing the advantages of ILMR in different applications, such as in sensing. Section V summarizes the paper.

## II. DESIGN AND THEORETICAL MODEL

The ILMR consists of a bus waveguide coupled to an infinity-shaped microresonator formed by two lobes joined together by a crossing (see Fig. 1). As shown in Fig. 1, both lobes are coupled to the bus waveguide. In addition, the crossing is ideal: the optical mode can only travel through it in a straight line and no excitation of the mode in the cross waveguide is possible (i.e., zero insertion losses and zero cross-talk). In the following, we assume that all the waveguides are single-mode. The relevant parameters of the ILMR are the resonance frequency ($\omega_0$), the two coupling rates between the lobes and the bus waveguide ($\Gamma_L$ and $\Gamma_R$), the total loss rate ($\gamma_{tot} = \gamma + \Gamma_L + \Gamma_R$), and the phase acquired by the optical mode that propagates in the bus waveguide between the two lobes of the infinity-shaped microresonator ($\varphi$). Note that we use the index R, L with reference to left and right in Fig. 1. The coupling coefficients $\Gamma_L$ and $\Gamma_R$ make the system geometrically symmetric if they assume the same value, or geometrically asymmetric if they assume different values.

As in a ring microresonator, the ILMR supports two counterpropagating optical modes: $\alpha_1$ and $\alpha_2$. The former, $\alpha_1$, is characterized by an electromagnetic wave, which, whenever it transits near the coupling region with the bus waveguide, is directed toward the center of the microresonator, namely, toward the crossing. While the latter mode, $\alpha_2$, travels through the ILMR in the opposite direction, see Fig. 1. As a result, these two modes interact differently with the bus waveguide. First of all, when an optical signal is coupled to the bus waveguide, $\alpha_1$ is excited first and, then, $\alpha_2$. This happens independently of the excitation direction (i.e., whether the signal is input from the left bus waveguide edge or the right bus waveguide edge). More importantly, only $\alpha_1$ can transfer energy to $\alpha_2$ by means of the bus waveguide. When $\alpha_1$ couples to the bus waveguide it propagates to the other coupling region and excites $\alpha_2$. On the contrary, when $\alpha_2$ couples to the bus waveguide, it can only propagate outward from the microresonator. Consequently, the coupling coefficient ($\beta_{12}$) that links $\alpha_1$ to $\alpha_2$ is different from zero ($\beta_{12} \neq 0$), while the coupling coefficient $\beta_{21}$ that links $\alpha_2$ to $\alpha_1$ is 0. Consequently, we can immediately claim that this simple structure works at an EP because it exhibits a completely asymmetric coupling between the two modes ($\beta_{12} \neq 0$ and $\beta_{21} = 0$).

In order to demonstrate rigorously that the ILMR works at an EP, we model an ideal ILMR, i.e., we neglect any waveguide surface-wall roughness that causes backscattering couples the two counterpropagating modes. By assuming that the ILMR works in the linear regime, i.e., low input optical intensities, $|E_{in}|^2 \ll 1$ mW, and $\omega_0 \gg \gamma_{tot}$, we can use the Temporal Coupled Mode Theory (TCMT)[49,50] to describe the system as

$$i\frac{d}{dt}\begin{pmatrix}\alpha_2 \\ \alpha_1\end{pmatrix} = \begin{pmatrix}\omega_0 - i\gamma_{tot} & -i\beta_{12} \\ 0 & \omega_0 - i\gamma_{tot}\end{pmatrix}\begin{pmatrix}\alpha_2 \\ \alpha_1\end{pmatrix}$$
$$-\begin{pmatrix}\sqrt{2\Gamma_R}e^{i\varphi} & \sqrt{2\Gamma_L}e^{i\varphi} \\ \sqrt{2\Gamma_L} & \sqrt{2\Gamma_R}\end{pmatrix}\begin{pmatrix}E_{in,L} \\ E_{in,R}\end{pmatrix}, \quad (1)$$

where the coupling between $\alpha_1$ and $\alpha_2$ reduces to $\beta_{12} = 4e^{i\varphi}\sqrt{\Gamma_L \Gamma_R}$. An alternative way to model this system is to use the Transfer Matrix Method (TMM); see Appendix A. Equation (1) shows that the Hamiltonian of the system is non-Hermitian and, in particular, is equivalent to that of other systems working at an EP, such as the Taiji microresonator.[25,36,44] Moreover, the eigenvalues and eigenvectors of the ILMR coalesce

$$\lambda_1 = \lambda_2 = \omega_0 - i\gamma_{tot}, \qquad v_1 = v_2 = \begin{pmatrix}1 \\ 0\end{pmatrix}. \quad (2)$$

Noteworthy, Eqs. (1) and (2) state that the ILMR is, respectively, a non-Hermitian system and works at an EP. Our device operates on an exceptional surface.[51,52] The ILMR operates at EP due to its geometrical shape. Thus, it remains at EP even if the characteristic parameters of the system (such as the resonant frequency, the propagation losses, and the coupling coefficients) are changed.

The output electric fields ($E_{out,R}, E_{out,L}$) are related to the input ones ($E_{in,L}, E_{in,R}$) and the internal modes of the ILMR in the following way:

$$\begin{pmatrix} E_{out,R} \\ E_{out,L} \end{pmatrix} = e^{i\varphi} \begin{pmatrix} E_{in,L} \\ E_{in,R} \end{pmatrix} + i \begin{pmatrix} \sqrt{2\Gamma_R} & \sqrt{2\Gamma_L} e^{i\varphi} \\ \sqrt{2\Gamma_L} & \sqrt{2\Gamma_R} e^{i\varphi} \end{pmatrix} \begin{pmatrix} \alpha_2 \\ \alpha_1 \end{pmatrix}. \quad (3)$$

Solving Eqs. (1) and (3) in the steady state and assuming $E_{in,L} := \varepsilon_{in,L} e^{-i\omega t}$, $E_{in,R} := \varepsilon_{in,R} e^{i\phi} e^{-i\omega t}$, and $\alpha_{1/2} := a_{1/2} e^{-i\omega t}$, we obtain

$$\varepsilon_{out,R} = e^{i\varphi}\left(1 - \frac{2(\Gamma_L + \Gamma_R)}{\gamma_{tot} - i\Delta\omega} + \frac{8\Gamma_L\Gamma_R}{(\gamma_{tot} - i\Delta\omega)^2}\right)\varepsilon_{in,L}$$
$$- \frac{4e^{i\varphi}\sqrt{\Gamma_L\Gamma_R}}{\gamma_{tot} - i\Delta\omega}\left(1 - \frac{2\Gamma_R}{\gamma_{tot} - i\Delta\omega}\right)\varepsilon_{in,R} e^{i\phi}, \quad (4)$$

$$\varepsilon_{out,L} = e^{i\varphi}\left(1 - \frac{2(\Gamma_L + \Gamma_R)}{\gamma_{tot} - i\Delta\omega} + \frac{8\Gamma_L\Gamma_R}{(\gamma_{tot} - i\Delta\omega)^2}\right)\varepsilon_{in,R} e^{i\phi}$$
$$- \frac{4e^{i\varphi}\sqrt{\Gamma_L\Gamma_R}}{\gamma_{tot} - i\Delta\omega}\left(1 - \frac{2\Gamma_L}{\gamma_{tot} - i\Delta\omega}\right)\varepsilon_{in,L}, \quad (5)$$

where $\Delta\omega = \omega - \omega_0$. Equations (4) and (5) can also be used in an interferometric excitation,[44,53] where the system is simultaneously excited from both the input ports with two coherent electromagnetic fields out of phase by $\phi$. To derive the formulas for a single side excitation, it is sufficient to impose $\varepsilon_{in,R}$ or $\varepsilon_{in,L}$ equal to zero. Equations (4) and (5) show that the intensities of the output fields do not depend on the phase acquired in the bus waveguide between the two lobes of the ILMR; the term $e^{i\varphi}$ can be collected. This means that it is possible to change the phase of the coefficient $\beta_{12} = 4e^{i\varphi}\sqrt{\Gamma_L\Gamma_R}$ without changing the response of the ideal ILMR.

Figure 2 reports the single excitation response of four different designed ILMR: (a) symmetric ILMR and (b)–(d) asymmetric ILMR. Here, with symmetric and asymmetric, we mean ILMR with the same coupling coefficients ($\Gamma_L = \Gamma_R$) or with two different coupling coefficients ($\Gamma_L \neq \Gamma_R$), respectively. The used parameters are reported in Table I. These parameters are those that were derived through simulations of the couplings during the design of the ILMRs. In the following, we identify the field intensity at the j output when the device is excited from i input as $|\varepsilon_{ij}|^2$, where i = L, R and j = L, R (L stands for left and R stands for right). Figure 2(a) reports counterintuitive transmissions and reflections. Since the ILMR works at an EP, one would expect that the transmission and reflection spectra would be characterized by a single dip and peak (coincident eigenvalues), respectively. Here, however, we do observe a doublet characterized by two dips and peaks having the same extinction rate. Note that this splitting is not due to the spurious coupling caused by backscattering,[46] but it is due to the interference of the different electric fields at the output; see Eqs. (4) and (5). It also emphasizes that the positions of minima in transmission are not the eigenvalues of the cavity under analysis.

By varying the coupling rates between the bus waveguide and the lobes of the ILMR ($\Gamma_L$ and $\Gamma_R$), one can obtain different spectral responses. Particularly, in Fig. 2(b), the transmission spectrum shows a quasi-negligible splitting, as well as in the reflection from the left side of the bus waveguide ($|\varepsilon_{LL}|^2$). Differently, the reflection from the opposite side ($|\varepsilon_{RR}|^2$) exhibits a clear doublet. Moreover, in resonance, the two reflections are completely different reaching a ratio of $|\varepsilon_{LL}|^2/|\varepsilon_{RR}|^2 \simeq 15$. By further varying the parameters $\Gamma_L$ and $\Gamma_R$ asymmetrically, one finds the spectral response reported in Fig. 2(c). Here, the reflection $|\varepsilon_{LL}|^2$, at the resonance frequency ($\Delta\omega = 0$), reduces to zero. This different spectral response demonstrates the potential of the ILMR. Interestingly, the relation between the parameters, which yield this spectral response, is easily calculated via Eq. (4). Imposing $|\varepsilon_{in,R}|^2 = 0$ and $|\varepsilon_{out,L}|^2 = 0$, at $\Delta\omega = 0$, gives $2\Gamma_L = \gamma_{tot}$ and then $\Gamma_L = \Gamma_R + \gamma$. Through the increase in $\Gamma_L$ and $\Gamma_R$ of the ILMR, it is also possible to exchange the transmission spectrum with the reflection spectrum by obtaining a peak in transmission and a dip in reflection at resonance [see Fig. 2(d)]. Noteworthy, these different spectral responses are obtained while remaining on an exceptional surface.

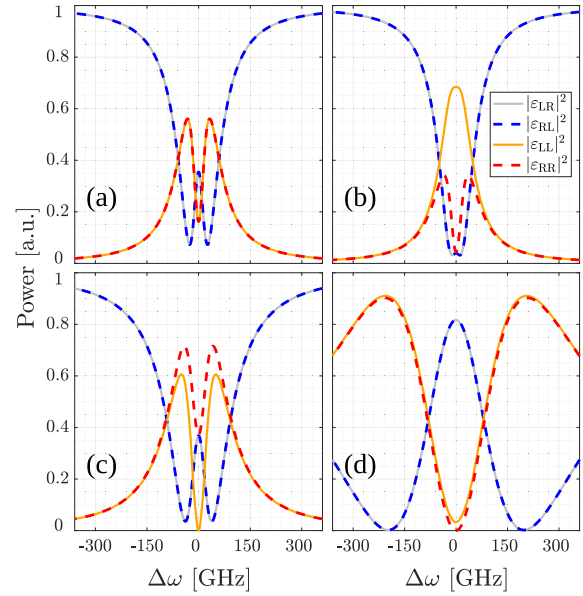

**FIG. 2.** Spectral response of four different ILMR: (a) a symmetric ILMR characterized by the same coupling coefficients ($\Gamma_L = \Gamma_R$) and (b)–(d) an asymmetric ILMR with different coupling coefficients ($\Gamma_L \neq \Gamma_R$). Solid lines identify the spectral responses of the system when it is excited from the left side of the bus waveguide, while dashed lines identify those related to excitation from the right. Gray and blue lines refer to transmission spectra, while orange and red lines refer to reflection spectra. The used parameters are the nominal ones for the fabricated ILMRs and are reported in Table I.

### A. Backscattering and sensing

An integrated microresonator having high-quality factor exhibits spurious backscattering mainly due to the surface roughness of the waveguides. The presence of this backscattering causes coupling between the counterpropagating modes of the microresonator, in our case $\alpha_1$ and $\alpha_2$. To include this phenomenon in our theory, we need to add to the model of the ideal ILMR Eq. (1) the following term:

$$\begin{pmatrix} 0 & -i\beta_{Bs,12} \\ -i\beta_{Bs,21} & 0 \end{pmatrix} \begin{pmatrix} \alpha_2 \\ \alpha_1 \end{pmatrix}, \quad (6)$$

where $\beta_{Bs,12}$ and $\beta_{Bs,21}$ are the backscattering coefficients that induce the spurious coupling between $\alpha_1$ and $\alpha_2$. Consequently, the Hamiltonian of the system results to

$$\mathcal{H} = \begin{pmatrix} \omega_0 - i\gamma_{tot} & -i(\beta_{12} + \beta_{Bs,12}) \\ -i\beta_{Bs,21} & \omega_0 - i\gamma_{tot} \end{pmatrix}. \quad (7)$$

**TABLE I.** Parameters used in Fig. 2.

| | $\Gamma_L$ [GHz] | $\Gamma_R$ [GHz] | $\gamma_{tot}$ [GHz] |
|---|---|---|---|
| Figure 2(a) | 12.6 | 12.6 | 35 |
| Figure 2(b) | 5.1 | 21.6 | 36.5 |
| Figure 2(c) | 25 | 15.2 | 50 |
| Figure 2(d) | 94.1 | 104 | 207.9 |

As a result, the eigenvalues and eigenvectors no longer coalesce, but are

$$\lambda_{1/2} = \omega_0 \pm i\sqrt{(\beta_{12} + \beta_{Bs,12})\beta_{Bs,21}} - i\gamma_{tot} \Rightarrow \lambda_1 \neq \lambda_2, \quad (8)$$

$$v_{1/2} = \frac{1}{\sqrt{\left|\frac{\beta_{12}+\beta_{Bs,12}}{\beta_{Bs,21}}\right| + 1}} \begin{pmatrix} \mp\sqrt{\frac{\beta_{12} + \beta_{Bs,12}}{\beta_{Bs,21}}} \\ 1 \end{pmatrix} \Rightarrow v_1 \nparallel v_2. \quad (9)$$

Usually backscattering, related, for example, to surface roughness, can be considered as a Hermitian perturbation to the ideal Hamiltonian, where $\delta\beta := \beta_{Bs,12} = -\beta_{Bs,21}^*$. Consequently, the splitting between the eigenvalues results in

$$\Delta\lambda = |\lambda_2 - \lambda_1| = \left|2i\sqrt{-(\beta_{12}\delta\beta^* + |\delta\beta|^2)}\right|. \quad (10)$$

Generalizing, we can consider the $\delta\beta$ perturbation not only caused by the surface-wall roughness but also due to any molecules/substances in the waveguide cladding. Therefore, $\delta\beta$ can simply be interpreted as the perturbation that we want to probe with an ILMR based optical sensor.

To determine the sensing potential of the ILMR toward a $\delta\beta$ perturbation, we assume that ideally both the real and imaginary parts of the eigenvalues can be measured. Figures 3(a) and 3(b) show the computed real and the imaginary parts of the eigenvalues divided by $|\beta_{12}|$ as a function of $\Re[\delta\beta/\beta_{12}]$ and $\Im[\delta\beta/\beta_{12}]$. Here, the two eigenvalues are shown with the red and blue surfaces (Riemann sheets[2,3,48]). Figure 3(a) shows that the real part of the two eigenvalues are equal when $\Im[\delta\beta/\beta_{12}] = 0$ and $-1 \leq \Re[\delta\beta/\beta_{12}] \leq 0$, at the same time Fig. 3(b) shows that the imaginary part of the two eigenvalues is equal when $\Im[\delta\beta/\beta_{12}] = 0$ and $\Re[\delta\beta/\beta_{12}] \leq -1 \vee \Re[\delta\beta/\beta_{12}] \geq 0$. In addition, Fig. 3(c) shows that only at (0, 0) and at (−1, 0) the inner product of the two eigenvectors, $|\langle v_1|v_2\rangle|$, is equal to one. As a result, starting from the ideal ILMR at (0, 0), we can reach another EP by varying the perturbation $\delta\beta$ in order to reach the point (−1, 0). Moreover, Fig. 3 shows that only the Riemann sheets in (a) are intersected, while in (b) they are not.

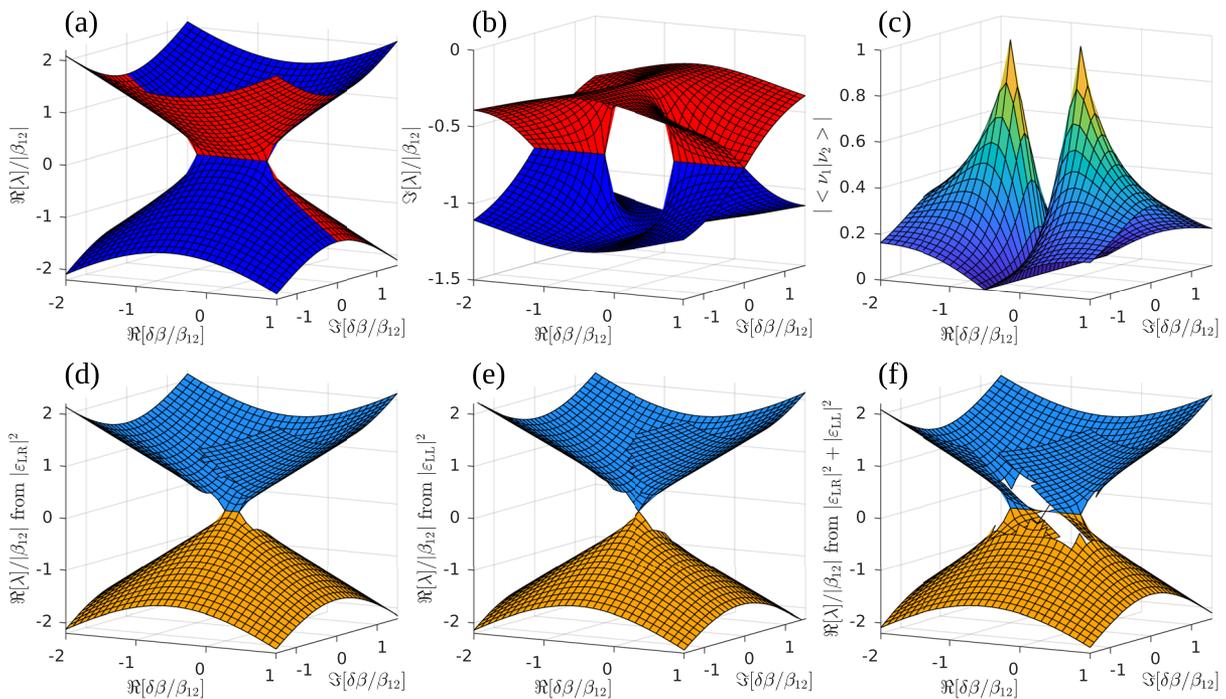

**FIG. 3.** Normalized Riemann sheets for the ILMR as a function of the real $\Re[\delta\beta]$ and the imaginary $\Im[\delta\beta]$ part of the Hermitian backscattering perturbation ($\delta\beta$). All the quantities are normalized to $\beta_{12}$. (a) and (b) report the maps of the real $\Re[\lambda]$ and of the imaginary $\Im[\lambda]$ parts of the eigenvalues. The red and blue surfaces refer to the first ($\lambda_1$) and the second ($\lambda_2$) eigenvalues, respectively. (c) shows the map of the absolute value of the inner product between the two eigenvectors ($|\langle v_1|v_2\rangle|$). (d)–(f) represent the map of the positions of the spectra minima for (d) and (f) and maxima for (e) of the transmission ($|\varepsilon_{LR}|^2$), reflection ($|\varepsilon_{LL}|^2$) and the sum of the two ($|\varepsilon_{LR}|^2 + |\varepsilon_{LL}|^2$), respectively. In orange the position of the dip with lower angular frequency and in Dodger blue the position of the dip with higher angular frequency. The holes in panels (d)–(f) are due to the inability to recognize two peaks. Here, we use $\Gamma_L = \Gamma_R = 1/4$ [a.u.] to get $\beta_{12} = 1$ [a.u.] and $\gamma = \Gamma_L$. Division by $\beta_{12}$ or its modulus was done to make the graphs independent of the relative phase between $\delta\beta$ and $\beta_{12}$ and of the absolute value of $\beta_{12}$.

As already discussed, the coefficient $\beta_{12}$ is equal to $4e^{i\varphi}\sqrt{\Gamma_L\Gamma_R}$, where $e^{i\varphi}$ is the acquired phase that the mode acquires by going from one lobe to the other through the bus waveguide. This means that by varying $\varphi$, for example, by means of a phase shifter component integrated into the bus waveguide, one can vary the eigenvalues of the system along a circular-like path with respect to the EP within the same perturbation $\delta\beta$. This encircling of the EP follows the Riemann sheets given in Fig. 3. In this way, we would be able to align $\beta_{12}$ with $\delta\beta$, thus being able to vary the spectral responsivity of the ILMR to this perturbation. Consequently, the variation of $\varphi$ increases the difference between the real part of the eigenvalues by causing $\Im[\delta\beta/\beta_{12}] = 0 \wedge \Re[\delta\beta/\beta_{12}] > 0$. With the ILMR, it is easy to implement a variation of the projection of the eigenvalues between the real and imaginary parts, but it is difficult to achieve a dynamic encircling of an EP.[8,10] The ILMR maintains the memory of its state only within the cavity coherence time (few ps).

To derive the real part $\Re[\lambda]$ of the eigenvalues, from the spectral responses of the system, it is customary to compute the spectral position of the two minima of the Autler–Townes splitting doublet.[21,54] However, as we already pointed out, the frequencies of the minima of the dips in the transmission spectrum do not give the eigenvalues. At an EP, the eigenvalues are equal, while, as shown in Fig. 2(a), the transmission and reflection of an ILMR exhibit a resonant doublet. Therefore, to use the ILMR as a sensor, let us consider a symmetric ILMR having $\Gamma_L = \Gamma_R = 1/4$ [a.u.] and $\gamma = \Gamma_{L/R}$. In this case, the normalized spectral minima or maxima for the transmitted power ($|\varepsilon_{LR}|^2$), the reflected power ($|\varepsilon_{LL}|^2$), and the sum of the two ($|\varepsilon_{LR}|^2 + |\varepsilon_{LL}|^2$) are shown in Figs. 3(d)–3(f), respectively. The spectrum of $|\varepsilon_{LR}|^2 + |\varepsilon_{LL}|^2$ yields the frequencies where the system has greater losses, which interestingly have a good correspondence with the eigenvalues of the system, as can be seen by comparing Figs. 3(a) and 3(f). In Figs. 3(d)–3(f), there are regions in which the Riemann sheets are not defined and, therefore, show a hole. In these regions, there is either only one non-zero peak in the spectrum or the second peak is masked by the first one. Finally, by simultaneously being able to measure the splitting relative to all the three spectra, and also by being able to encircle an EP to satisfy $\Im[\delta\beta/\beta_{12}] = 0 \wedge \Re[\delta\beta/\beta_{12}] \geq 0$, it is possible to improve the accuracy in the $\delta\beta$ perturbation sensing.

This last conclusion is evidenced by looking at the splitting ($\Re[\Delta\lambda]/|\beta_{12}|$) as a function of $\Re[\delta\beta/\beta_{12}]$ by imposing $\Im[\delta\beta/\beta_{12}] = 0$ (Fig. 4). By comparing Figs. 4(a) and 4(b), we observe that the doublet is replaced by a single dip in the spectrum of the sum of the transmission and reflection of the ideal ILMR. The spectrum of $|\varepsilon_{LR}|^2 + |\varepsilon_{LL}|^2$ as a function of $\delta\beta$ is shown in Fig. 10 in Appendix E, and the analytic expressions of the spectral minima are reported in Appendix D. Furthermore, Fig. 4(c) shows that the actual splitting of the eigenvalues $\Re[\Delta\lambda]$ is well represented by the splitting in the spectrum of $|\varepsilon_{LR}|^2 + |\varepsilon_{LL}|^2$, as also shown in Figs. 3(a) and 3(f). As can also be derived from Eq. (10), for small perturbations $\delta\beta$, the splitting follows the square-root characteristic of EPs.[21,22,25] Figure 4(c) shows that, for $\delta\beta$ around and greater than zero, both the splittings in the transmission spectra and the reflection spectra vary linearly as a function of the perturbation. Therefore, by using the ILMR one can simultaneously use both the enhanced sensing of the EP due to the square root dependence as well as the linear dependence, which can be very useful during sensor calibration. Moreover, we observe that the latter enlarges the working region of the sensor by covering the part near $\delta\beta = 0$.

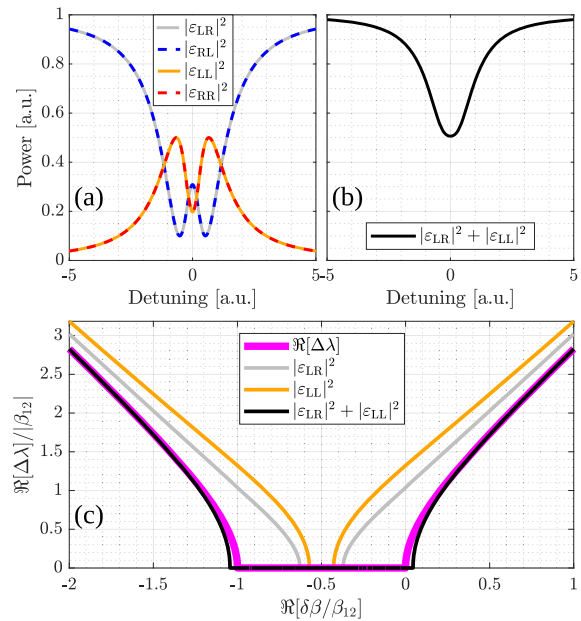

**FIG. 4.** Output spectra of a symmetric ILMR ($\Gamma_L = \Gamma_R$). (a) shows the different transmission and reflection spectra for left or right excitation. (b) reports the sum of the output field intensities at the right (transmission) and left (reflection) ports when the ILMR is excited from left ($|\varepsilon_{LR}|^2 + |\varepsilon_{LL}|^2$). (c) represents with different colors the splitting of the eigenvalues or the doublets observed in the spectra of $|\varepsilon_{LR}|^2$, $|\varepsilon_{LL}|^2$ and $|\varepsilon_{LR}|^2 + |\varepsilon_{LL}|^2$ as a function of the backscattering perturbation $\delta\beta/\beta_{12}$. Here, we use $\Gamma_L = \Gamma_R = 1/4$ [a.u.] to get $\beta_{12} = 1$ [a.u.] and $\gamma = \Gamma_L$.

By fixing the conditions for coherent perfect absorption (CPA) in a symmetric ILMR,[34,45,55] a peculiar output spectrum is obtained. The CPA condition is determined by fixing one of the eigenvalues of the scattering matrix of the symmetric ILMR equal to zero (see Appendix C). If we consider real frequencies, the CPA condition is realized with $\gamma_{tot} = 2(\Gamma_L + \Gamma_R)$ and the resulting spectra are given in Fig. 5, where we used $\Gamma_L = \Gamma_R = \Gamma$, $\Im[\delta\beta] = 0$, $\Im[\beta_{12}] = 0$, $\Re[\delta\beta] \geq 0$, and $\Re[\beta_{12}] \geq 0$. The transmission and reflection spectra are different and show doublet dips and peaks [Fig. 5(a)]. In contrast, the spectrum of $|\varepsilon_{LR}|^2 + |\varepsilon_{LL}|^2$ [black curve in Fig. 5(b)] has the typical quartic pattern.[34,45,55] However, the plateau is 0.5 a.u.. and it is not zero for perfect absorption. In our case, coherent perfect absorption is manifested only by an interferometric excitation,[44,55] i.e., by exciting the ILMR through both bus waveguide edges. In this way, the plateau of the quartic response goes to zero, as shown by the dark green curve in Fig. 5(b) that represents the right output field intensity. As expected, this transmission overlaps with the spectral dependence of the first eigenvalue of the scattering matrix given in Appendix C and plotted in Fig. 5(b) with a yellow line. Interestingly, under the CPA condition, the actual splitting of the eigenvalues [magenta curve in Fig. 5(c)] equals the splitting of the spectrum of $|\varepsilon_{LR}|^2 + |\varepsilon_{LL}|^2$ (black curve). Consequently, there is no region of null splitting around the zero of the perturbation. At the CPA EP, the splitting evaluated from the spectrum of $|\varepsilon_{LR}|^2 + |\varepsilon_{LL}|^2$ depends as $2\sqrt{\delta\beta(\delta\beta + 4\Gamma)}$. This corresponds to the splitting of the eigenvalues of the Hamiltonian reported in Eq. (10).

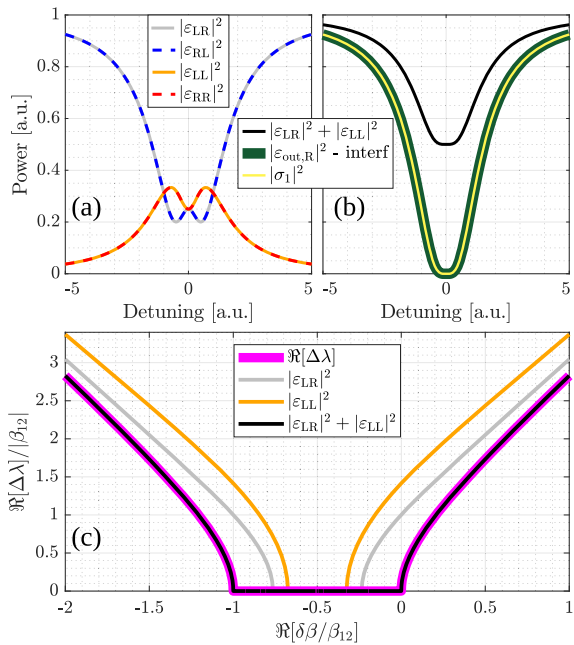

**FIG. 5.** Output spectra of a symmetric ILMR ($\Gamma_L = \Gamma_R$) in the coherent perfect absorption condition $\gamma_{tot} = 2(\Gamma_L + \Gamma_R)$. (a) Transmission and reflection spectra for left or right excitation. (b) The sum of the output field intensities at the right (transmission) and left (reflection) ports when the ILMR is excited from left ($|\varepsilon_{LR}|^2 + |\varepsilon_{LL}|^2$, black curve). The dark green line reports $|\varepsilon_{out,R}|^2$ when a symmetric interferometric excitation[44] ($|\varepsilon_{in,L}|^2 = |\varepsilon_{in,R}|^2 = 1$) is used. The yellow curve shows the square absolute value of the first eigenvalue of the scattering matrix of the ILMR ($|\sigma_1|^2$). (c) Splitting of the eigenvalues or the doublets observed in the spectra of $|\varepsilon_{LR}|^2$, $|\varepsilon_{LL}|^2$, and $|\varepsilon_{LR}|^2 + |\varepsilon_{LL}|^2$ as a function of the backscattering perturbation $\delta\beta/\beta_{12}$. The color code is given in the inset. Here we use $\Gamma_L = \Gamma_R = 1/4$ [a.u.] to get $\beta_{12} = 1$ [a.u.] and $\gamma = \Gamma_L + \Gamma_R = 1/2$ [a.u.] to reach the coherent perfect absorption condition.

## III. EXPERIMENTAL MEASUREMENTS

### A. Experimental setup and samples

To verify the theoretical prediction we have designed four different ILMRs: one symmetric ($\Gamma_L = \Gamma_R$) and the others asymmetric ($\Gamma_L \neq \Gamma_R$); see Fig. 6. These have been fabricated with $450 \times 220$ nm$^2$ cross section silicon waveguides embedded in a silica cladding by the IMEC/Europractice facility within a multi-project wafer program.

The symmetrical ILMR geometry is characterized by a crossing, four $3\pi/4$ Euler curves with a minimum radius of 15 $\mu$m, point coupling regions, and straight waveguides to connect the various elements and draw the bus waveguide. The gap widths between the loops and the bus waveguide are 165 nm. The crossings used are those of the design kit of the IMEC/Europractice facility, which from our measurements at 1550 nm have losses of $0.18 \pm 0.01$ dB/crossing and negligible reflections. The asymmetric ILMR geometry has straight coupling regions (length $\simeq 10$ $\mu$m) and gaps that differ for the various asymmetric geometries gap$_L$ = [392, 288, 205] nm and gap$_R$ = [300, 326, 198] nm. In addition, the loops are achieved with Euler curves with a minimum radius equal to 10 $\mu$m.

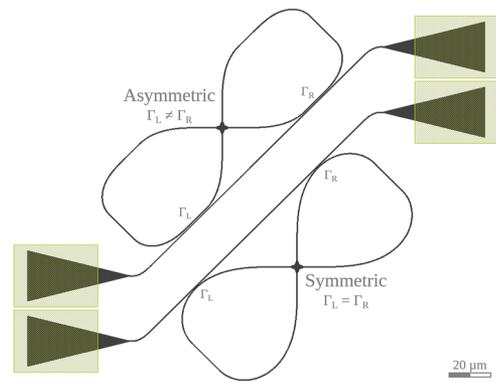

**FIG. 6.** The design of an asymmetric (top) and a symmetric (bottom) ILMRs. The silicon waveguides are indicated by the black lines. The yellow boxes identify the input/output grating regions.

The ILMRs were measured by using an interferometric optical setup.[44] It allows simultaneous measurement of transmission and reflection in both the excitation directions. Briefly, the set-up is based on a Continuous-Wave tunable laser whose output beam is divided by a fiber splitter into two arms. These arms contain a Variable Optical Attenuator (VOA), a fiber polarization controller, and an optical circulator with one of its outputs connected to an InGaAs photodetector. Both arms end on a stripped fiber that couples light to the sample. We used a laser wavelength of around 1.55 $\mu$m and a power of about 10 $\mu$W.

### B. Results

The spectral response of the four different ILMRs was measured for both excitation directions (Fig. 7). The transmission and reflection spectra from both sides are plotted with gray/blue and orange/red lines, respectively. In particular, the gray and orange lines correspond to an input excitation from the left while the blue and red lines from the right. The dashed black lines refer to the fit of the spectra by the theory of Sec. II, which was performed on the four spectral responses with the same parameters (fit parameters are reported in Table II).

As expected, four different types of spectral responses are observed in Fig. 7. Spectra are similar to those reported in Fig. 2. Their asymmetry with respect to a detuning $\Delta\omega = 0$ is caused by the backscattering induced by the surface-wall roughness. Indeed, the theoretical fit which includes the backscattering matches the experimental data. This validates the modeling given by the system of equations [Eqs. (1), (3), (6), and (7)], reported in Sec. II. The fit parameters reported in Table II show that the backscattering is non-Hermitian because $n := (\beta_{Bs,12} + \beta^*_{Bs,21})/2 \neq 0$.[44,46] As a further check, if we let $|\beta_{12}|$ run as a free parameter in the fit, we obtain values that are compatible with $4\sqrt{\Gamma_L \Gamma_R}$. This also remarks on the validity of our theoretical model.

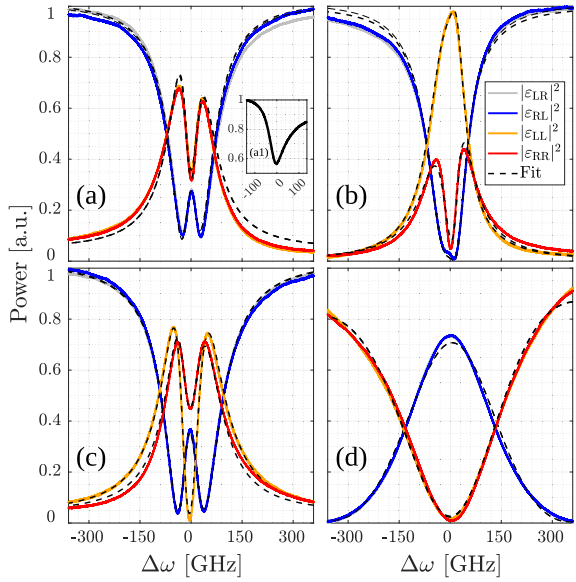

**FIG. 7.** Experimental spectra of four different ILMRs. The dashed black lines identify the fits with the temporal coupled mode theory equations. The solid lines represent the experimental data. In gray and blue lines are reported the transmission spectra while in orange and red lines are reported the reflection spectra. Moreover, the gray and orange lines correspond to an input excitation from the left while the blue and red lines from the right. The inset (a1) shows the sum of the transmitted and reflected intensities, when the ILMR is excited from left ($|\varepsilon_{LR}|^2 + |\varepsilon_{LL}|^2$).

Figure 7 (a1, inset) shows the sum of the two measured output intensities ($|\varepsilon_{LR}|^2 + |\varepsilon_{LL}|^2$) for a left excitation of the symmetric ILMR. In this spectrum, as expected from the theoretical simulations (Fig. 4), the splitting disappears. Only one minimum is observed at approximately zero detuning (resonance). The asymmetry of the peak is attributed to the presence of the backscattering, which is a non-Hermitian perturbation whose complex vector [$\beta_{Bs,12}$ in Eq. (7)] does not have the same direction as $\beta_{12}$; see Table II.

## IV. DISCUSSION

In the literature, other microresonator geometries that work at an EP have been demonstrated.[21,25,34,36,45,51,52] The symmetric ILMR differs from all these devices because, although the eigenvalues and eigenvectors coalesce, the spectral responses in transmission and reflection exhibit resonance splitting. Even the devices described in Refs. 36, 51, and 52, while having the same Hamiltonian as the ILMR, do not have the same spectral response because their scattering matrix is different from that of the ILMR.

It is interesting to compare in detail the ILMR and the Taiji geometry[36] since they are topological equivalent.[56] Assuming ideal crossings, the ILMR is equivalent to a Taiji microresonator excited by the S-shaped waveguide; see Fig. 9(b) in Appendix A. However, while in the case of a Taiji microresonator the bus waveguide is not necessary for the resonator to work at an exceptional point, for the ILMR the bus waveguide is essential. This means that the total losses of the Taiji microresonator, having further couplings of the bus waveguide for input and output, are higher than those of the ILMR, see Fig. 8(b). Moreover, it is easy to see that the ILMR is also equivalent to the structure schematized in Fig. 9(c) in Appendix A, which is characterized by a microring resonator having two couplings with the bus waveguide, and the crossing no longer turns out to be internal to the microring resonator, but between the two couplings in the bus waveguide. This ILMR-equivalent structure may appear similar to the device reported in Ref. 52. However, the latter differs from the ILMR by the presence of an optical isolator that halves the coupling from $\alpha_1$ to $\alpha_2$. Thus, unlike the ILMR, the response of this device does not exhibit a doublet in the transmission spectrum when its response is not perturbed by the presence of a scatterer.

**TABLE II.** Parameters derived through fits of experimental data reported in Fig. 7.

|  | Figure 7(a) | Figure 7(b) | Figure 7(c) | Figure 7(d) |
| --- | --- | --- | --- | --- |
| $\Gamma_L$ [GHz] | 12.86(7) | 5.81(3) | 23.90(6) | 183(1) |
| $\Gamma_R$ [GHz] | 13.25(7) | 30.8(3) | 12.1(1) | 182(1) |
| $\gamma_{tot}$ [GHz] | 39.9(3) | 43.7(3) | 47.0(3) | 400(2) |
| $|\beta_{12}|$ [GHz] | 52.2(8) | 54(1) | 68(1) | 732(14) |
| $\arg[\beta_{12}]$ | 0.1436(6) | −2.148(3) | −1.865(6) | 1.64(1) |
| $|\beta_{Bs,12}|$ [GHz] | 2.21(2) | 5.90(3) | 5.93(6) | 18.77(1) |
| $\arg[\beta_{Bs,12}]$ | 0.782(7) | 1.567(8) | 1.85(1) | 1.000(9) |
| $|\beta_{Bs,21}|$ [GHz] | 3.40(2) | 13.1(1) | 8.90(8) | 23.12(1) |
| $\arg[\beta_{Bs,21}]$ | 1.11(1) | 0.981(4) | 1.047(7) | 1.77(1) |

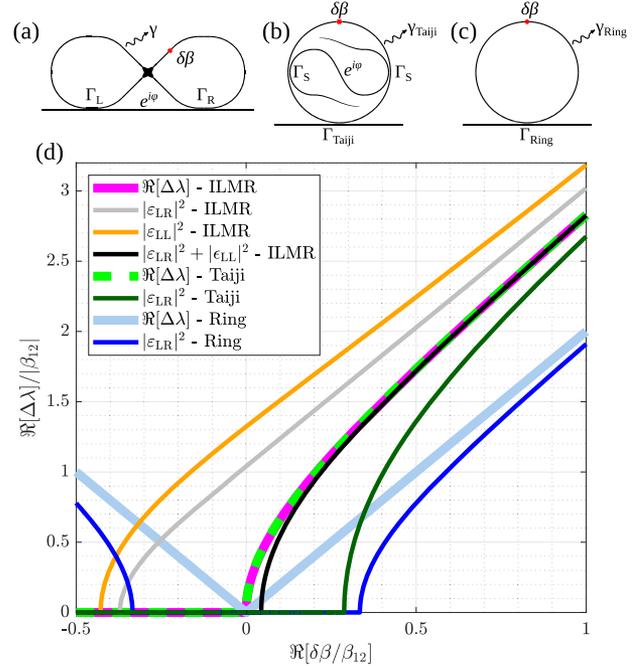

**FIG. 8.** Splitting of the eigenvalues or doublets present in the spectra as a function of the perturbation $\delta\beta/\beta_{12}$ for three different structures: an ILMR (a), a Taiji microresonator (b), and a microring resonator (c). Here, we used $\Gamma_{Ring} = \Gamma_{Taiji} = \Gamma_S = \Gamma_L = \Gamma_R = 1/4$ [a.u.] to get $\beta_{12} = 1$ [a.u.], $\beta_{12,Taiji} = 4e^{i\varphi}\Gamma_S = 1$ [a.u.], and $\gamma_{Ring} = \gamma_{Taiji} = \gamma = \Gamma_L$. In this figure, the conditions $\Im[\delta\beta/\beta_{12}] = 0$ and $\delta\beta = \beta_{Bs,12} = -\beta^*_{Bs,21}$ were used. The thick lines represent eigenvalues, while the thin lines represent experimentally measurable quantities.

Let us further compare the sensitivity to an external perturbation of the ILMR with those of other types of integrated microresonators. In Fig. 8, we show the ILMR (a), the Taiji microresonator (b), and the simple microring resonator (c). Here, we assume the same propagation losses and that the coupling coefficients are equal to the loss rate ($\gamma_{\text{Ring}} = \gamma_{\text{Taiji}} = \gamma = \Gamma_L$, $\Gamma_{\text{Ring}} = \Gamma_{\text{Taiji}} = \Gamma_S = \Gamma_L = \Gamma_R = 1/4$ [a.u.], $\beta_{12} = 4e^{i\varphi}\sqrt{\Gamma_L \Gamma_R} = 1$ [a.u.], and $\beta_{12,\text{Taiji}} = 4e^{i\varphi}\Gamma_S = 1$ [a.u.]). The equations used are given in Appendix B. Note that with these parameters, the microring resonator is in the most favorable coupling regime, the critical coupling one. Figure 8 shows that the eigenvalue splittings for the ILMR and the Taiji microresonator follow the expected square-root dependence on small perturbations. The splitting for an EP microresonator is greater than that for a structure working at a diabolic point, such as the simple microresonator. In this last case, a linear trend in splitting vs the perturbation is observed.[21,25] Although the trend of the eigenvalues is very interesting, it is not easy to access them directly in an experiment; the spectral positions of the transmission minima do not accurately reflect the eigenvalues. Therefore, in order to make a more useful comparison from an application point of view, we have reported the transmission minima splittings obtained from the spectra in Fig. 8.

From the experimentally observable transmission spectra, there is always a region where two distinct minima cannot be observed because the spectral width of the peaks makes them indistinguishable or the noise masks the splitting.[11,12,21] However, by using the ILMR, one has the possibility of using not only the sum of the transmitted and the reflected intensity, i.e., the square-root trend, but also the two intensities separately, i.e., the linear trend where the splitting is maintained. By exploiting this characteristic feature of the ILMR, one is also able to use the region around small perturbations where the other structures are unusable because of resolution or noise. Furthermore, it is observed that using $|\varepsilon_{LR}|^2 + |\varepsilon_{LL}|^2$ in an ILMR one can decrease the width of this region; see Fig. 8. In addition, one can vary the phase $\varphi$ between the two lobes of the ILMR or in the S-shaped waveguide for the Taiji microresonator to align $\beta_{12}$ with the perturbation $\delta\beta$, and thus one achieves a maximum splitting in the optical response. Such a condition corresponds to rotate $\beta_{12}$ so that $\Re[\delta\beta/\beta_{12}] \geq 0$ and $\Im[\delta\beta/\beta_{12}] = 0$. This alignment within the complex plane is easily performed with the ILMR to be compared with the difficulty of tuning nano-tips as in the case of Ref. 21. The responsivity and sensitivity of the ILMR to backscattering perturbations compared with those of the Taiji microresonator and a simple microring resonator are higher, especially for small perturbations, see Appendix F. In addition, by imposing the coherent perfect absorption condition for the ILMR [$\gamma_{\text{tot}} = 2(\Gamma_L + \Gamma_R)$], we can obtain that the spectral minima of $|\varepsilon_{LR}|^2 + |\varepsilon_{LL}|^2$ faithfully follow the values of the eigenvalues, thus improving the sensing performance of the ILMR. A last remark is about the role of the crossing in the ILMR, which is one critical element of the proposed geometry. If crossing losses are predominant over other losses, the efficiency in distinguishing small perturbations with the square root trend would decrease; however, the linear trend still provides a way to detect the perturbation. This is another advantage of the ILMR that makes it robust with respect to imperfections.

## V. CONCLUSIONS

We have proposed a novel integrated photonic microresonator that works on an exceptional surface: the Infinity-Loop Microresonator (ILMR). This device, consisting of an infinity-shaped waveguide coupled with a bus waveguide at its two lobes, has several interesting features. The ILMR is always at an Exceptional Point (EP), whatever the two coupling coefficients with the bus waveguide. A symmetric ILMR presents a doublet in the spectral responses. By varying the couplings, different shapes in transmission and reflection with absolute zero reflection or zero transmission at resonance can be obtained. By different side excitation, one can either get identical reflections or completely different reflections. In addition, the condition for obtaining perfect absorption for the ILMR was derived. Under this condition, and when the ILMR is geometrically symmetric, there is a quartic behavior as a function of detuning in the absorption spectrum of the system. All these features are achieved while keeping the ILMR at an EP.

Furthermore, we have studied the improvements in sensing applications provided by the ILMR, which are related to the possibility to observe at the same time both a square-root (sensitivity enhancement) and a linear dependence of the spectral splitting as a function of a Hermitian perturbation. This feature washes out the insensitive region of an EP optical sensor caused by resolution or noise. This last observation strengthens the potential of the ILMR as a device that could exploit both the characteristics of an EP and the linearity of a diabolic point.

The geometrical asymmetry and the ease of control of the coupling phase between the two lobes via integrated phase shifters make the ILMR a suitable tunable element for more complex geometries where arrays of interconnected ILMR can be used to realize integrated topological structures or to study the EP from a quantum point of view.

## ACKNOWLEDGMENTS

We gratefully thank Dr. Iacopo Carusotto for useful inputs and valuable comments. This work was supported by Q@TN, the joint lab between the University of Trento, FBK—Fondazione Bruno Kessler, INFN—National Institute for Nuclear Physics, and CNR—National Research Council. S.B. acknowledges the co-financing of the European Union FSE-REACT-EU, PON Research and Innovation 2014–2020 DM1062/2021.

We acknowledge funding from PAT through the Q@TN joint lab, from Ministero dell'Istruzione, dell'Università e della Ricerca [PRIN PELM (20177 PSCKT)] and European Union FSE-REACT-EU, PON Research and Innovation 2014–2020 DM1062/2021.

## AUTHOR DECLARATIONS
### Conflict of Interest
The authors have no conflicts to disclose.

### Author Contributions
**Riccardo Franchi**: Conceptualization (equal); Data curation (lead); Formal analysis (lead); Investigation (lead); Methodology (equal); Software (equal); Writing – original draft (lead). **Stefano Biasi**: Conceptualization (equal); Investigation (equal); Methodology (equal); Supervision (equal); Writing – review & editing (equal). **Diego Piciocchi**: Data curation (equal); Formal analysis (supporting); Investigation (supporting); Software (equal); Writing – review & editing (supporting). **Lorenzo Pavesi**: Funding acquisition (equal); Investigation (equal); Project administration (lead); Supervision (lead); Writing – review & editing (equal).

## DATA AVAILABILITY
Data underlying the results presented in this paper are not publicly available at this time but may be obtained from the authors upon reasonable request.

## APPENDIX A: TRANSFER MATRIX METHOD

As reported in the main text, the three structures sketched in Fig. 9 are equivalent, as demonstrated by the fact that their optical modes follow the same equations.

In order to describe the ILMR, the Transfer Matrix Method (TMM) can also be used. Using the coefficients given in Fig. 9, where $t$ are the transmission coefficients of the specific coupling region while $k$ are the coupling coefficients, we are able to write the following system of equations:

$$
\begin{aligned}
E_1 &= t_1 E_0 + ik_1 E_9, & E_{0r} &= t_1 E_{1r} + ik_1 E_{4r}, \\
E_2 &= e^{i\varphi} E_1, & E_{1r} &= e^{i\varphi} E_{2r}, \\
E_3 &= t_2 E_2 + ik_2 E_{8r}, & E_{2r} &= t_2 E_{3r} + ik_2 E_7, \\
E_4 &= t_1 E_9 + ik_1 E_0, & E_{4r} &= e^{i\psi_{45}} E_{5r}, \\
E_5 &= e^{i\psi_{45}} E_4, & E_{5r} &= t_{Bs} E_{6r} - b_{Bs,12} E_5, \\
E_6 &= t_{Bs} E_5 - b_{Bs,21} E_{6r}, & E_{6r} &= e^{i\psi_{67}} E_{7r}, \\
E_7 &= e^{i\psi_{67}} E_6, & E_{7r} &= t_2 E_{8r} + ik_2 E_2, \\
E_8 &= t_2 E_7 + ik_2 E_{3r}, & E_{8r} &= e^{i\psi_{89}} E_{9r}, \\
E_9 &= e^{i\psi_{89}} E_8, & E_{9r} &= t_1 E_{4r} + ik_1 E_{1r}, \\
\psi_{jl} &= \frac{2\pi}{\lambda} n_{eff} L_{jl} + i\alpha L_{jl}, & L &= L_{45} + L_{67} + L_{89}, \\
\psi &= \frac{2\pi}{\lambda} n_{eff} L + i\alpha L.
\end{aligned}
\quad (A1)
$$

In the following, for simplicity, we assumed that in the couplings between the bus waveguide and the lobes there are no losses ($t_{1/2}^2 + k_{1/2}^2 = 1$). The parameter $L_{ij}$ identifies the waveguide length between the numbers $i$ and $j$ reported in Fig. 9 $\varphi$, as in the main text, is the acquired phase to go from 1 to 2; see Fig. 9. We associated the backscattering with a scatterer (black star in Fig. 9) characterized by the coefficients $t_{Bs}$, $-b_{Bs,12}$, and $-b_{Bs,21}$. The losses due to the crossing are absorbed either in the propagation losses ($\alpha$) or in the factor $t_{Bs}$, while the possible coupling between the counter-propagating modes given by crossing can be incorporated within the coefficients $b_{Bs,12}$ and $b_{Bs,21}$. To be noted that this crossing cross-talk turns out to be negligible compared with the backscattering given by the surface roughness of the waveguides. From Eq. (A1), we derive

$$\varepsilon_{LR} = \varepsilon_{RL} = \frac{N}{D}, \quad \varepsilon_{LL} = \frac{N_{LL}}{D}, \quad \varepsilon_{RR} = \frac{N_{RR}}{D}, \quad (A2)$$

$$\varepsilon_{out,R} = \varepsilon_{in,L}\varepsilon_{LR} + \varepsilon_{in,R}e^{i\phi}\varepsilon_{RR}, \quad (A3)$$

$$\varepsilon_{out,L} = \varepsilon_{in,R}e^{i\phi}\varepsilon_{RL} + \varepsilon_{in,L}\varepsilon_{LL}, \quad (A4)$$

where

$$N = \left[(t_2^2 - 2)t_1^2 - 2t_2^2 + 1\right] t_{Bs} e^{i(\psi+\varphi)}$$
$$+ t_1 \left(k_1 k_2 e^{i(2\psi_{45}+\psi_{89})} b_{Bs,12} + t_2 e^{i\varphi}\right) t_1 e^{i(2\psi_{67}+\psi_{89}+\varphi)}$$
$$\times \left[t_2 e^{i(2\psi_{45}+\psi_{89})}\left(t_{Bs}^2 - b_{Bs,12} b_{Bs,21}\right) - k_1 k_2 e^{i\varphi} b_{Bs,21}\right], \quad (A5)$$

$$N_{LL} = 2k_1 k_2 e^{i(\psi_{45}+\psi_{67}+\varphi)}\left[-t_1 t_{Bs} + t_2 e^{i\psi}\left(t_{Bs}^2 - b_{Bs,12} b_{Bs,21}\right)\right]$$
$$- (t_1^2 - 1)e^{2i\psi_{45}} b_{Bs,12} + (t_2^2 - 1)t_1^2 \left(-e^{2i(\psi_{67}+\varphi)}\right) b_{Bs,21}, \quad (A6)$$

$$N_{RR} = e^{2i\psi_{89}}\left[(t_2^2-1)t_1^2\left(-e^{2i\psi_{45}}\right)b_{Bs,12} - (t_1^2-1)e^{2i(\psi_{67}+\varphi)}b_{Bs,21}\right]$$
$$+ 2k_1 k_2 e^{i(\psi_{89}+\varphi)}\left(-t_2 + t_1 e^{i\psi} t_{Bs}\right), \quad (A7)$$

$$D = 1 + t_2 e^{i(\psi_{67}+\psi_{89})}\left[t_2 t_1^2 e^{i(\psi_{45}+\psi)}\left(t_{Bs}^2 - b_{Bs,12} b_{Bs,21}\right)\right.$$
$$\left. - 2k_1 k_2 e^{i(\psi_{67}+\varphi)} b_{Bs,21} - 2t_1 e^{i\psi_{45}} t_{Bs}\right]. \quad (A8)$$

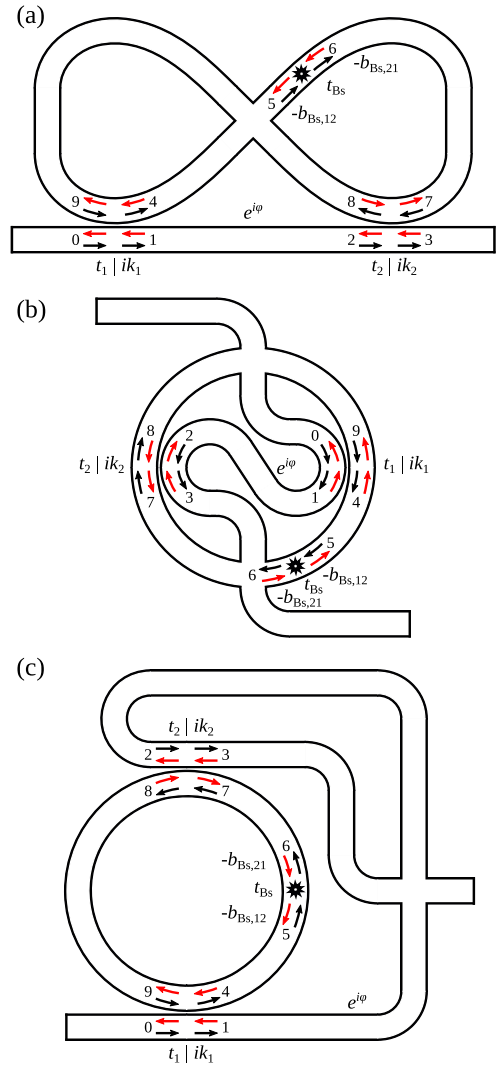

**FIG. 9.** Sketches of microresonators ideally equal to an ILMR. (a) ILMR, (b) Taiji microresonator excited from the S-shaped waveguide, and (c) microresonator having two couplings with the bus waveguide having within it a crossing. The black arrows identify the $E_n$ fields, with $n = 0, 1, \ldots, 9$, and their propagation direction. Instead, the red arrows identify the fields $E_{nr}$, having opposite propagation directions. All the parameters are described in the text.

Below, to simplify the discussion we impose the following conditions on the position of the scatterer inside the ILMR: $\psi_{89} = 2\psi_{45} = 2\psi_{67} = \psi/2$ ($L_{89} = 2L_{45} = 2L_{67} = L/2$).

If the conditions describing a situation of "low" couplings ($k_{1/2} \ll 1$, $t_{1/2} \simeq 1$, $|b_{Bs,12/21}| \ll 1$, $t_{Bs} \simeq 1$, $\sigma \simeq 1$) are met, and performing the following substitutions:

$$t_{1/2} \simeq 1 - \frac{\Gamma_{L/R}}{\tilde{f}}, \quad k_{1/2} \simeq \sqrt{\frac{2\Gamma_{L/R}}{\tilde{f}}}, \quad (A9)$$

$$t_{Bs}\sigma \simeq 1 - \frac{\gamma}{\tilde{f}}, \quad \sigma := e^{-\alpha L}, \quad (A10)$$

$$\beta_{12} = 4e^{i\varphi}\sqrt{\Gamma_L \Gamma_R}, \quad (A11)$$

$$b_{Bs,12} \simeq \frac{\beta_{Bs,12}}{\tilde{f}}, \quad b_{Bs,21} \simeq \frac{\beta_{Bs,21}}{\tilde{f}}, \quad (A12)$$

$$\tilde{f} := \frac{c}{n_g L} = \frac{c\,\mathrm{FSR}}{\lambda_0^2}, \quad \mathrm{FSR} := \frac{\lambda_0^2}{n_g L}, \quad (A13)$$

$$\Re[\psi] = \frac{\Delta\omega}{\tilde{f}}, \quad (A14)$$

it is straightforward to verify that we get the same results derived through the Temporal Coupled Mode Theory (TCMT) equations reported in Eqs. (1), (3), (6), and (7), namely,

$$\varepsilon_{out,R} = \varepsilon_{in,L} e^{i\varphi}\left[1 + \frac{-2\Gamma_R(-i\Delta\omega+\gamma_{tot}) + 2\Gamma_L(-i\Delta\omega+\gamma_{tot}-4\Gamma_R)}{(-i\Delta\omega+\gamma_{tot})^2 - \left(\beta_{Bs,12}+4e^{i\varphi}\sqrt{\Gamma_L\Gamma_R}\right)\beta_{Bs,21}}\right.$$
$$+ \frac{2\sqrt{\Gamma_L\Gamma_R}\left(e^{-i\varphi}\beta_{Bs,12}+e^{i\varphi}\beta_{Bs,21}\right)}{(-i\Delta\omega+\gamma_{tot})^2 - \left(\beta_{Bs,12}+4e^{i\varphi}\sqrt{\Gamma_L\Gamma_R}\right)\beta_{Bs,21}}\right] + \varepsilon_{in,R} e^{i\phi} e^{i\varphi}$$
$$\times \left[-\frac{4\sqrt{\Gamma_L\Gamma_R}(-i\Delta\omega+\gamma_{tot}-2\Gamma_L)}{(-i\Delta\omega+\gamma_{tot})^2 - \left(\beta_{Bs,12}+4e^{i\varphi}\sqrt{\Gamma_L\Gamma_R}\right)\beta_{Bs,21}}\right.$$
$$\left.+ \frac{2e^{i\varphi}\Gamma_L\beta_{Bs,21}+2e^{-i\varphi}\Gamma_R\beta_{Bs,12}}{(-i\Delta\omega+\gamma_{tot})^2 - \left(\beta_{Bs,12}+4e^{i\varphi}\sqrt{\Gamma_L\Gamma_R}\right)\beta_{Bs,21}}\right], \quad (A15)$$

$$\varepsilon_{out,L} = \varepsilon_{in,R} e^{i\phi} e^{i\varphi}\left[1 + \frac{-2\Gamma_R(-i\Delta\omega+\gamma_{tot}) + 2\Gamma_L(-i\Delta\omega+\gamma_{tot}-4\Gamma_R)}{(-i\Delta\omega+\gamma_{tot})^2 - \left(\beta_{Bs,12}+4e^{i\varphi}\sqrt{\Gamma_L\Gamma_R}\right)\beta_{Bs,21}}\right.$$
$$+ \frac{2\sqrt{\Gamma_L\Gamma_R}\left(e^{-i\varphi}\beta_{Bs,12}+e^{i\varphi}\beta_{Bs,21}\right)}{(-i\Delta\omega+\gamma_{tot})^2 - \left(\beta_{Bs,12}+4e^{i\varphi}\sqrt{\Gamma_L\Gamma_R}\right)\beta_{Bs,21}}\right] + \varepsilon_{in,L} e^{i\varphi}$$
$$\times \left[-\frac{4\sqrt{\Gamma_L\Gamma_R}(-i\Delta\omega+\gamma_{tot}-2\Gamma_L)}{(-i\Delta\omega+\gamma_{tot})^2 - \left(\beta_{Bs,12}+4e^{i\varphi}\sqrt{\Gamma_L\Gamma_R}\right)\beta_{Bs,21}}\right.$$
$$\left.+ \frac{2e^{-i\varphi}\Gamma_L\beta_{Bs,12}+2e^{i\varphi}\Gamma_R\beta_{Bs,21}}{(-i\Delta\omega+\gamma_{tot})^2 - \left(\beta_{Bs,12}+4e^{i\varphi}\sqrt{\Gamma_L\Gamma_R}\right)\beta_{Bs,21}}\right]. \quad (A16)$$

In Eq. (A13), $\tilde{f}$ is one over the cavity round-trip time.

In Eqs. (A15) and (A16), it is observed that only if $\beta_{Bs,12} \neq 0$ or/and $\beta_{Bs,21} \neq 0$, the intensities depends on $\varphi$. Note that if one uses a symmetrical ILMR ($\Gamma_L = \Gamma_R$) and assumes a Hermitian perturbation ($\delta\beta = \beta_{Bs,12} = -\beta_{Bs,21}^*$), then the change of $\varphi$ (rotation in the complex plane of $\beta_{12}$) corresponds to a rotation of $\delta\beta$ in the negative direction ($e^{-i\varphi}$).

In conclusion, we have demonstrated that, if the following relations are satisfied ($k_{1/2} \ll 1$, $t_{1/2} \simeq 1$, $|b_{Bs,12/21}| \ll 1$, $t_{Bs}\sigma \simeq 1$) equivalent to ($\Gamma_L$ & $\Gamma_R$ & $\gamma_{tot}$ & $|\beta_{12}|$ & $|\beta_{Bs,12}|$ & $|\beta_{Bs,21}| \ll \omega_0$), to move from a TCMT to a TMMs model one has to use the following relations:

$$\Gamma_{L/R} \simeq \tilde{f}(1-t_{1/2}), \quad \Gamma_{L/R} \simeq \tilde{f} k_{1/2}^2/2, \quad (A17)$$

$$\gamma \simeq \tilde{f}(1-t_{Bs}\sigma), \quad \sigma := e^{-\alpha L}, \quad (A18)$$

$$\beta_{12} = 4e^{i\varphi}\sqrt{\Gamma_L\Gamma_R} \simeq 4e^{i\varphi}\tilde{f}\sqrt{(1-t_1)(1-t_2)}, \quad (A19)$$

$$\beta_{12} \simeq 2e^{i\varphi}\tilde{f} k_L k_R, \quad (A20)$$

$$\beta_{Bs,12} \simeq \tilde{f} b_{Bs,12}, \quad \beta_{Bs,21} \simeq \tilde{f} b_{Bs,21}, \quad (A21)$$

$$\tilde{f} := \frac{c}{n_g L} = \frac{c\,\mathrm{FSR}}{\lambda_0^2}, \quad \mathrm{FSR} := \frac{\lambda_0^2}{n_g L}. \quad (A22)$$

### APPENDIX B: TCMT FOR THE TAIJI AND THE RING MICRORESONATOR

To generate Fig. 8 in Sec. IV, we used the TCMT equations for the Taiji and the microring resonator described below.

The TCMT equations that describe the Taiji are[36,44]

$$i\frac{d}{dt}\begin{pmatrix}\alpha_{CCW}\\ \alpha_{CW}\end{pmatrix} = \begin{pmatrix}\omega_0 - i\gamma_{tot,Taiji} & -i(\beta_{12,Taiji}+\beta_{Bs,12})\\ -i\beta_{Bs,21} & \omega_0 - i\gamma_{tot,Taiji}\end{pmatrix}$$
$$\times \begin{pmatrix}\alpha_{CCW}\\ \alpha_{CW}\end{pmatrix} - \sqrt{2\Gamma_{Taiji}}\begin{pmatrix}E_{in,L}\\ E_{in,R}\end{pmatrix}, \quad (B1)$$

$$\begin{pmatrix}E_{out,R}\\ E_{out,L}\end{pmatrix} = \begin{pmatrix}E_{in,L}\\ E_{in,R}\end{pmatrix} + i\sqrt{2\Gamma_{Taiji}}\begin{pmatrix}\alpha_{CCW}\\ \alpha_{CW}\end{pmatrix}, \quad (B2)$$

where, $\gamma_{tot,Taiji} = \gamma_{Taiji} + \Gamma_{Taiji} + 2\Gamma_S$ is the total loss rate of the Taiji microresonator composed by the propagation/absorption/scattering losses, the coupling losses with the bus waveguide and with the S-shaped waveguide [see Fig. 8(b)]. Moreover, the coefficient $\beta_{12,Taiji} = 4e^{i\varphi}\Gamma_S$.

The TCMT equations that describe the Ring microresonator are[44]

$$i\frac{d}{dt}\begin{pmatrix}\alpha_{CCW}\\ \alpha_{CW}\end{pmatrix} = \begin{pmatrix}\omega_0 - i\gamma_{tot,Ring} & -i\beta_{Bs,12}\\ -i\beta_{Bs,21} & \omega_0 - i\gamma_{tot,Ring}\end{pmatrix}$$
$$\times \begin{pmatrix}\alpha_{CCW}\\ \alpha_{CW}\end{pmatrix} - \sqrt{2\Gamma_{Ring}}\begin{pmatrix}E_{in,L}\\ E_{in,R}\end{pmatrix}, \quad (B3)$$

$$\begin{pmatrix}E_{out,R}\\ E_{out,L}\end{pmatrix} = \begin{pmatrix}E_{in,L}\\ E_{in,R}\end{pmatrix} + i\sqrt{2\Gamma_{Ring}}\begin{pmatrix}\alpha_{CCW}\\ \alpha_{CW}\end{pmatrix}, \quad (B4)$$

where, $\gamma_{tot,Ring} = \gamma_{Ring} + \Gamma_{Ring}$ is the total loss rate of the ring microresonator composed by the propagation/absorption/scattering losses and the coupling losses with the bus waveguide [see Fig. 8(c)].

## APPENDIX C: SCATTERING MATRIX AND COHERENT PERFECT ABSORPTION CONDITION

The scattering matrix of the ideal ILMR is

$$S = \begin{pmatrix}\varepsilon_{RL} & \varepsilon_{LL}\\ \varepsilon_{RR} & \varepsilon_{LR}\end{pmatrix}. \quad (C1)$$

The eigenvalues of the scattering matrix are

$$\sigma_{1/2} = e^{i\varphi}\left(\frac{\Delta\omega - 2i\Gamma_R + i\gamma_{tot}}{\Delta\omega + i\gamma_{tot}} + \frac{2\Gamma_L(-i\Delta\omega - 4\Gamma_R + \gamma_{tot})}{(\Delta\omega + i\gamma_{tot})^2}\right.$$
$$\left.\pm \frac{4\sqrt{\Gamma_L\Gamma_R(i\Delta\omega + 2\Gamma_L - \gamma_{tot})(i\Delta\omega + 2\Gamma_R - \gamma_{tot})}}{(\Delta\omega + i\gamma_{tot})^2}\right). \quad (C2)$$

Only the first eigenvalue of the scattering matrix ($\sigma_1$) has a zero, while only the second eigenvalue ($\sigma_2$) has a pole. The zero of $\sigma_1$ is for

$$\omega_{\sigma_1,zero} = \omega_0 - i(\gamma_{tot} - 2(\Gamma_L + \Gamma_R)). \quad (C3)$$

Instead, the pole of $\sigma_2$ is for

$$\omega_{\sigma_2,pole} = \omega_0 - i\gamma_{tot}. \quad (C4)$$

Note that the pole of $\sigma_2$ corresponds to the eigenvalues of the Hamiltonian of the system [Eq. (8)]. The ideal ILMR being on an exceptional surface always shows an Hamiltonian characterized by degenerate eigenvalues.

The coherent perfect absorption (CPA) condition is obtained when an eigenvalue of the scattering matrix is zero, restricted to the case of real frequencies. This condition holds for $\gamma_{tot} = 2(\Gamma_L + \Gamma_R)$. Satisfying this condition $\omega_{\sigma_1,zero}$ is only real.[45,55]

Under the condition that the ILMR is geometrically symmetrical ($\Gamma_L = \Gamma_R = \Gamma$) and the condition of coherent perfect absorption is satisfied, $|\sigma_1|^2$ has a quartic trend around zero detuning ($\Delta\omega = 0$), as shown in Fig. 5(b) in the main text. Expanding in Taylor series $|\sigma_1|^2$ around $\Delta\omega = 0$, we obtain

$$|\sigma_1|^2 = \frac{\Delta\omega^4}{256\,\Gamma^4} - \frac{\Delta\omega^6}{2048\,\Gamma^6} + \frac{3\Delta\omega^8}{65\,536\,\Gamma^8} + O[\Delta\omega^{10}]. \quad (C5)$$

As a result, even the absorption line shape of the first eigenchannel has the quartic dependence as observed in Refs. 34, 45, and 55.

It is worth noting that, under these conditions, the spectrum of the sum between transmission and reflection ($|\varepsilon_{LR}|^2 + |\varepsilon_{LL}|^2$) also has a quartic dependence. Expanding in the Taylor series, we have

$$|\varepsilon_{LR}|^2 + |\varepsilon_{LL}|^2 = \frac{1}{2} + \frac{\Delta\omega^4}{512\,\Gamma^4} - \frac{\Delta\omega^6}{4096\,\Gamma^6} + O[\Delta\omega^8]. \quad (C6)$$

Note that, in this case, $|\sigma_1|^2 = 2(|\varepsilon_{LR}|^2 + |\varepsilon_{LL}|^2) - 1$. In addition, we also have $|\sigma_1|^2 = |\varepsilon_{LR} + \varepsilon_{RR}|^2$, as shown in Fig. 5(b).

## APPENDIX D: ANALYTIC EXPRESSION OF THE SPECTRAL MINIMA

In this section, the analytic functions of the spectral minima of $|\varepsilon_{LR}|^2 + |\varepsilon_{LL}|^2$ for a geometrically symmetric ILMR ($\Gamma_L = \Gamma_R = \Gamma$) are derived. To simplify the equations, we assume that $\beta_{Bs,12} = -\beta_{Bs,21}^* = \delta\beta$, $\Im[\delta\beta] = 0$, $\Im[\beta_{12}] = 0$, $\Re[\delta\beta] \geq 0$, and $\Re[\beta_{12}] \geq 0$. By imposing these conditions, we obtain

$$|\varepsilon_{LR}|^2 + |\varepsilon_{LL}|^2$$
$$= \frac{1}{(4\Gamma\delta\beta + \delta\beta^2 - \Delta\omega^2)^2 + 2\gamma_{tot}^2(4\Gamma\delta\beta + \delta\beta^2 + \Delta\omega^2) + \gamma_{tot}^4}.$$
$$\times \left((8\Gamma^2 + 4\Gamma(\delta\beta - \Delta\omega) + (\delta\beta - \Delta\omega)^2 - 4\Gamma\gamma_{tot} + \gamma_{tot}^2)\right.$$
$$\times (8\Gamma^2 + 4\Gamma(\delta\beta + \Delta\omega) + (\delta\beta + \Delta\omega)^2 - 4\Gamma\gamma_{tot} + \gamma_{tot}^2)$$
$$\left.+ 16\Gamma^2(\Delta\omega^2 + (\gamma_{tot} - 2\Gamma)^2)\right). \quad (D1)$$

Two spectral minima are observed only when the following condition is met:

$$2(2\Gamma + \delta\beta)\sqrt{(2\Gamma + \delta\beta)^2 - 4\Gamma\gamma_{tot} + \gamma_{tot}^2} + 4\Gamma\gamma_{tot}$$
$$\geq 8\Gamma^2 + 4\Gamma\delta\beta + \delta\beta^2 + \gamma_{tot}^2. \quad (D2)$$

When this relationship is satisfied, the spitting is

$$\Delta\lambda_{|\varepsilon_{LR}|^2+|\varepsilon_{LL}|^2} = 2\left|8\Gamma^2 + 4\delta\beta\Gamma - 4\gamma_{tot}\Gamma + \delta\beta^2\right.$$
$$+ \gamma_{tot}^2 - 2(2\Gamma + \delta\beta)\sqrt{(2\Gamma + \delta\beta)^2 + \gamma_{tot}^2 - 4\Gamma\gamma_{tot}}\Big|^{1/2}$$
$$\times \cos\left[\frac{1}{2}\arg\left[-8\Gamma^2 - 4\Gamma\delta\beta - \delta\beta^2\right.\right.$$
$$+ 2(2\Gamma + \delta\beta)\sqrt{(2\Gamma + \delta\beta)^2 - 4\Gamma\gamma_{tot} + \gamma_{tot}^2}$$
$$\left.\left.+ 4\Gamma\gamma_{tot} - \gamma_{tot}^2\right]\right]. \quad (D3)$$

With the parameters used in Fig. 4, where $\gamma_{tot} = 3\Gamma$, we have that the splitting is non-zero for

$$\delta\beta > \left(\sqrt{\frac{2\sqrt{13}+7}{3}} - 2\right)\Gamma. \quad \text{(D4)}$$

By writing the Taylor series of the splitting for $\delta\beta \to \tilde{\delta\beta} + \left(\sqrt{(2\sqrt{13}+7)/3} - 2\right)\Gamma$, we get

$$\Delta\lambda_{|\varepsilon_{LR}|^2+|\varepsilon_{LL}|^2} = 2^{3/4}\sqrt[4]{13(1+\sqrt{13})}\sqrt{\Gamma}\sqrt{\tilde{\delta\beta}} + O\left[\tilde{\delta\beta}^{3/2}\right]. \quad \text{(D5)}$$

Note the square root dependence of the splitting as a function of $\tilde{\delta\beta}$.

In the CPA case, where $\gamma_{tot} = 4\Gamma = \beta_{12}$, we have that the splitting is non-zero for $\delta\beta > 0$ and it is equal to

$$\Delta\lambda_{|\varepsilon_{LR}|^2+|\varepsilon_{LL}|^2} = 2\sqrt{\delta\beta(4\Gamma+\delta\beta)}. \quad \text{(D6)}$$

It is worth noting that the splitting derived by calculating the position of the spectral minima of $|\varepsilon_{LR}|^2 + |\varepsilon_{LL}|^2$ in the CPA condition is equivalent to the splitting of the eigenvalues of the Hamiltonian Eq. (10), as shown in Fig. 5(c).

## APPENDIX E: SPECTRA AS A FUNCTION OF THE BACKSCATTERING PERTURBATION

Figure 10 shows the spectra of $(|\varepsilon_{LR}|^2 + |\varepsilon_{LL}|^2)$ as a function of $\delta\beta$. Note that $1 - (|\varepsilon_{LR}|^2 + |\varepsilon_{LL}|^2)$ is the system absorption. Panel (a) shows the case characterized by the parameters used in Fig. 4, while panel (b) shows the case corresponding to the CPA condition used in Fig. 5.

In both panels, it can be seen that for $\delta\beta = 0$ there is no doublet, while increasing the perturbation shows the increase of the splitting.

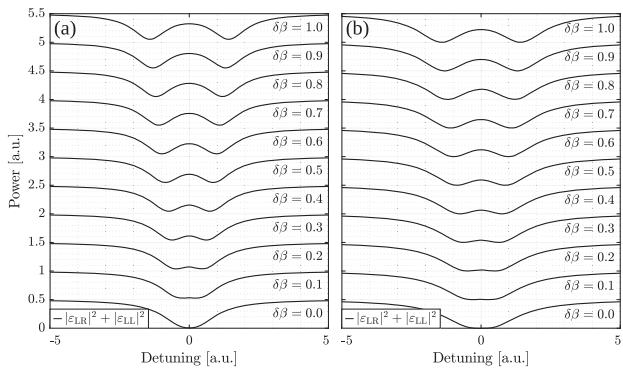

**FIG. 10.** Spectra of the sum of the output field intensities at the right (transmission) and left (reflection) ports when the ILMR is excited from left as a function of the backscattering perturbation ($\delta\beta$). The spectra are vertically shifted according to $|\varepsilon_{LR}|^2 + |\varepsilon_{LL}|^2 + (n/2 - 1)$, where $n = 1, 2, \ldots, 11$ identifies the different perturbations indicated on the right of each spectrum. Panel (a) shows the case of Fig. 4 in which $\Gamma_L = \Gamma_R = \Gamma = 1/4$ [a.u.] to get $\beta_{12} = 1$ [a.u.] and $\gamma = \Gamma$, and thus $\gamma_{tot} = 3\Gamma$. Panel (b) shows the case of Fig. 5 in which the condition to obtain the coherent perfect absorption is satisfied. In panel (b), we use $\Gamma_L = \Gamma_R = \Gamma = 1/4$ [a.u.] to get $\beta_{12} = 1$ [a.u.] and $\gamma = \Gamma_L + \Gamma_R = 1/2$ [a.u.], and thus $\gamma_{tot} = 4\Gamma$.

Comparing panels (a) and (b) of Fig. 10, we can see that in panel (b) the extinction ratio of the two dips remains constant at 2, while in panel (a) it decreases with increasing $\delta\beta$. Furthermore, at $\delta\beta = 0$ in panel (b), we observe a quartic dependence on detuning that is not observed in (a).

## APPENDIX F: COMPARISON OF THE THREE MICRORESONATOR STRUCTURES IN TERMS OF RESPONSIVITY AND SENSITIVITY

Let us compare the sensing efficiency of an ILMR with a simple microring resonator (MR) and a Taiji microresonator (TJMR) having the same propagation losses. We define the responsivity enhancement as $\Re[\Delta\lambda]/\Re[\Delta\lambda_{MR}]$, i.e., the responsivity with respect to the ideal splitting of the eigenvalues of the microresonator. Figure 11 shows the responsivity enhancement as a function of the perturbation. We see that for $\Re[\delta\beta/\beta_{12}] \simeq 0.1$, using the spectrum $|\varepsilon_{LR}|^2 + |\varepsilon_{LL}|^2$ of the ILMR yields a responsivity that is 2.5 times higher than that of the MR eigenvalues. This result is in agreement with the enhancement factor obtained experimentally in Ref. 21. Furthermore, using the spectra $|\varepsilon_{LR}|^2$ and $|\varepsilon_{LL}|^2$ separately, we observe a 2.5 times larger splitting for $\Re[\delta\beta/\beta_{12}]$ values below about 0.4. If we look at the splitting curves obtained from the MR or the TJMR transmission spectra, we observe a lower responsivity than for ILMR with the same propagation losses.

Figure 12 shows the relative sensitivity of the ILMR with respect to the sensitivity computed by the splitting of the MR eigenvalues. The sensitivity is computed by the derivative of the splitting as a function of the backscattering perturbation $\delta\beta$. Figure 12 shows that the ILMR has a higher sensitivity than the MR, which is about 2.5 times higher for $\Re[\delta\beta/\beta_{12}] \simeq 0.1$ (black curve, $|\varepsilon_{LR}|^2 + |\varepsilon_{LL}|^2$). Using the spectra of $|\varepsilon_{LR}|^2$ and $|\varepsilon_{LL}|^2$ separately, the sensitivity is almost constant and almost equal to 1 even for very small perturbations, whereas the sensitivity obtained using the MR or TJMR spectra is zero at very small perturbations since the splitting vanishes; see Fig. 8.

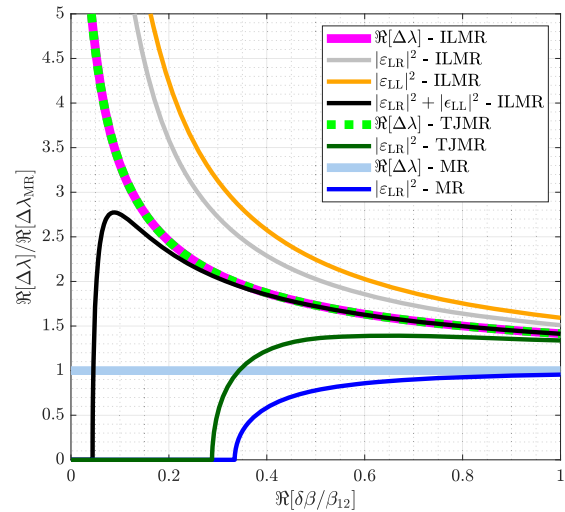

**FIG. 11.** Ratio of the splittings shown in Fig. 8 with the splitting of the microring resonator eigenvalues as a function of the backscattering perturbation strength $\delta\beta$. Here, we used $\Gamma_{Ring} = \Gamma_{Taiji} = \Gamma_S = \Gamma_L = \Gamma_R = 1/4$ [a.u.] to get $\beta_{12} = 1$ [a.u.], $\beta_{12,Taiji} = 4e^{i\varphi}\Gamma_S = 1$ [a.u.], and $\gamma_{Ring} = \gamma_{Taiji} = \gamma = \Gamma_L$. In this figure, the conditions $\Im[\delta\beta/\beta_{12}] = 0$ and $\delta\beta = \beta_{Bs,12} = -\beta_{Bs,21}^*$ were used. The thick lines represent eigenvalues, while the thin lines represent experimentally measurable quantities.

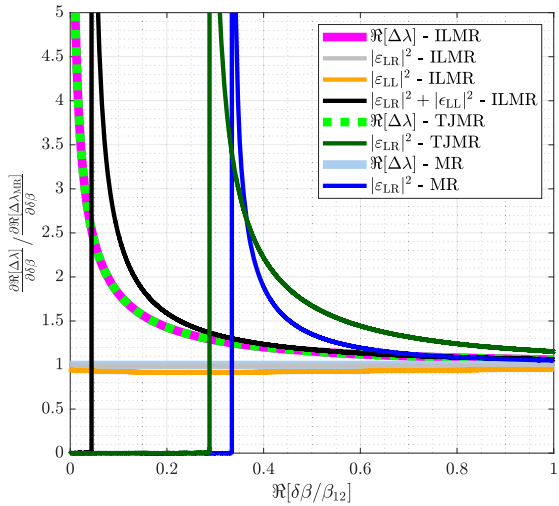

**FIG. 12.** Ratio of the derivative of the splittings shown in Fig. 8 to that corresponding to the splitting of the microring resonator eigenvalues as a function of the backscattering perturbation strength $\delta\beta$. Here, we used $\Gamma_{Ring} = \Gamma_{Taiji} = \Gamma_S = \Gamma_L = \Gamma_R = 1/4$ [a.u.] to get $\beta_{12} = 1$ [a.u.], $\beta_{12,Taiji} = 4e^{i\varphi}\Gamma_S = 1$ [a.u.], and $\gamma_{Ring} = \gamma_{Taiji} = \gamma = \Gamma_L$. In this figure, the conditions $\Im[\delta\beta/\beta_{12}] = 0$ and $\delta\beta = \beta_{Bs,12} = -\beta^*_{Bs,21}$ were used. The thick lines represent eigenvalues, while the thin lines represent experimentally measurable quantities.

Although the ILMR has both higher responsivity and a larger sensitivity at small perturbations than the MR and the TJMR, in this paper, we did not investigate how the signal-to-noise ratio behaves in the three different structures.[12] The latter may penalize the ILMR more than a simple MR, partially reducing the advantage of using an ILMR. This aspect needs a dedicated study that is beyond the scope of this paper.